\documentclass{aa}
\usepackage{graphicx}
\usepackage{subfigure}
\usepackage{txfonts}
\begin{document}
   \title{Synchrotron flaring in the jet of 3C~279}
   
   %\subtitle{}
         \author{E.J. Lindfors
          \inst{1,2}, 
          M. T\"urler
	  \inst{3,4},
          E. Valtaoja\inst{1,5},
	  H. Aller\inst{6},
	  M. Aller\inst{6},
	  D. Mazin\inst{7},
	  C. M. Raiteri\inst{8},
	  J. A. Stevens\inst{9},
	  M. Tornikoski\inst{2},
	  G. Tosti\inst{10},
	  \and 
	  M. Villata\inst{8}\fnmsep
          }

   \offprints{E.J. Lindfors email:elilin@utu.fi}

   \institute{Tuorla Observatory, V\"ais\"al\"a Institute of Space Physics and Astronomy, University of Turku, 21500 Piikki\"o, Finland
         \and
             Mets\"ahovi Radio Observatory, Helsinki University of Technology, 02540 Kylm\"al\"a, Finland
          \and
             Geneva Observatory, ch. des Maillettes 51, 1290 Sauverny, Switzerland
         \and 
             INTEGRAL Science Data Centre, ch. d'Ecogia 16, CH-1290 Versoix, Switzerland   
         \and 
             Department of Physics, University of Turku, 20100 Turku, Finland
         \and
	     Astronomy Department, University of Michigan, Ann Arbor, MI
         48109, United States
	 \and
	     Max-Planck-Institut f\"ur Physik, M\"unchen, Germany
	 \and
	     INAF-Osservatorio Astronomico di Torino, Via Osservatorio 20,
         I-10025 Pino Torinese, Italy
	 \and
	     Centre for Astrophysics Research, Science and Technology Research
         Institute, University of Hertfordshire, College Lane, Herts, AL10 9AB
	 \and
	     Osservatorio Astronomico di Perugia, via Bonfigli, 06123 Perugia,
         Italy}
   \date{Received, accepted}

   \abstract
{}
{We study the synchrotron flaring behaviour of the blazar
3C~279 based on an extensive dataset covering 10 years of monitoring
at 19 different frequencies in the radio-to-optical range.}
{The
properties of a typical outburst are derived from the observations by
decomposing the 19 lightcurves into a series of self-similar
events. This analysis is achieved by fitting all data simultaneously to
a succession of outbursts defined according to the shock-in-jet model of
Marscher \& Gear (1985).} 
{We compare the derived properties of the
synchrotron outbursts in 3C~279 to those obtained with a similar
method for the quasar 3C~273. It is argued that differences in the flaring
behaviour of these two sources are intrinsic to the sources themselves rather
than being due to orientation effects. 
We also compare the start times and flux densities of our
modelled outbursts with those measured from radio components identified
in Very Long Baseline Interferometry (VLBI) images. We find
VLBI counterparts for most of our model outbursts, although some
high-frequency peaking events are not seen in the radio maps. Finally,
we study the link between the appearance of a new synchrotron component and the EGRET
gamma-ray state of the source at 10 different epochs. We find that
an early-stage shock component is always present during high gamma-ray states,
while in low gamma-ray states the time since the onset of the last
synchrotron outburst is significantly longer. This statistically
significant correlation supports the idea that gamma-ray flares are
associated with the early stages of shock components propagating in the
jet. We note, however, that the shock wave is already beyond the
broad line region during the gamma-ray flaring.}
{}

   \titlerunning{Synchrotron flaring in the jet of 3C~279}
   \authorrunning{E. J. Lindfors et al.}
   \maketitle
%
%________________________________________________________________

\section{Introduction}
It is generally accepted that radio outbursts in active galactic nuclei (AGN)
are triggered by growing shocks in a relativistic jet.  There have been many
efforts to identify the individual components in radio to submillimeter (submm)
lightcurves. Litchfield et al. (1995) and Stevens et al. (1995, 1996, 1998)
managed to follow the early evolution of individual outbursts by subtracting
the quiescent contribution (assumed to be constant) from the total
spectrum. Valtaoja et al. (1999) identified outbursts by decomposing the
variations in millimeter (mm) and centimeter lightcurves into exponential
flares. T\"urler et al. (1999) measured the properties of synchrotron outbursts
by decomposing the radio-to-submillimeter wavelength lightcurves into a series
of self-similar flaring events.  The method was further developed by T\"urler
et al.  (2000) who used the shock model of Marscher \& Gear (1985) to describe
both the average evolution of the outbursts and their individual
characteristics. This model was found to provide a good description of the
lightcurves of 3C~273. In this work, we decompose the lightcurves of a second
source, 3C~279. Since our adopted methodology is analogous to that used by
T\"urler et al. (2000) the results are directly comparable to those found for
3C~273.

While in previous studies the outbursts were identified in the
radio-to-submillimeter regime, we here follow the synchrotron spectrum
up to infrared and optical frequencies. Searches for correlations in the
radio-to-optical emission from AGN have been conducted since the 1970s.
One of the most recent works was presented in Hanski et
al. (2002). In their sample of 20 AGNs they found a clear radio-to-optical
correlation in seven cases, a possible correlation in six cases and
no correlation for the remainder. This agrees well with the findings of
previous studies: there appears to be some kind of correlation between the radio and
optical emission but this correlation is not seen in all sources at all epochs. The
general trend seems to be that all radio outbursts are accompanied by optical
outbursts but conversely not all optical outbursts have radio counterparts.

The source 3C~279 is one of the brightest and most variable blazars at radio
wavelengths. As such, it is monitored regularly in all radio wavebands from the
centimeter to the submillimeter domain. In the optical, the historical
lightcurve shows variations with a typical amplitude of about 2 magnitudes, but
reaching 8 magnitudes during flares (Webb et al. 1990). The weak blue bump of
3C~279 allows us to follow the synchrotron spectrum up to infrared and optical
wavelengths. Indeed, 3C~279 is one of the few objects for which a clear
correlation is found between the radio and optical wavebands (Tornikoski et
al. 1994).

In this paper we present results of a multifrequency lightcurve decomposition
of 3C~279. The method used is based on T\"urler et al. (1999; 2000) with some
modifications for the specific case of 3C~279, including an extension of the
analysis to infrared and optical wavelengths. The average evolution of
outbursts is compared to that of 3C~273. We also compare our results with (1)
VLBI maps from Wehrle et al. (2001) and Jorstad et al. (2004) to study the
connection between the model outbursts and VLBI components, and (2) gamma-ray
data obtained by Hartman et al. (2001) to study the connection between the
synchrotron spectra and the gamma-ray state.

\begin{figure*}
\centering \includegraphics[angle=0,width=17cm]{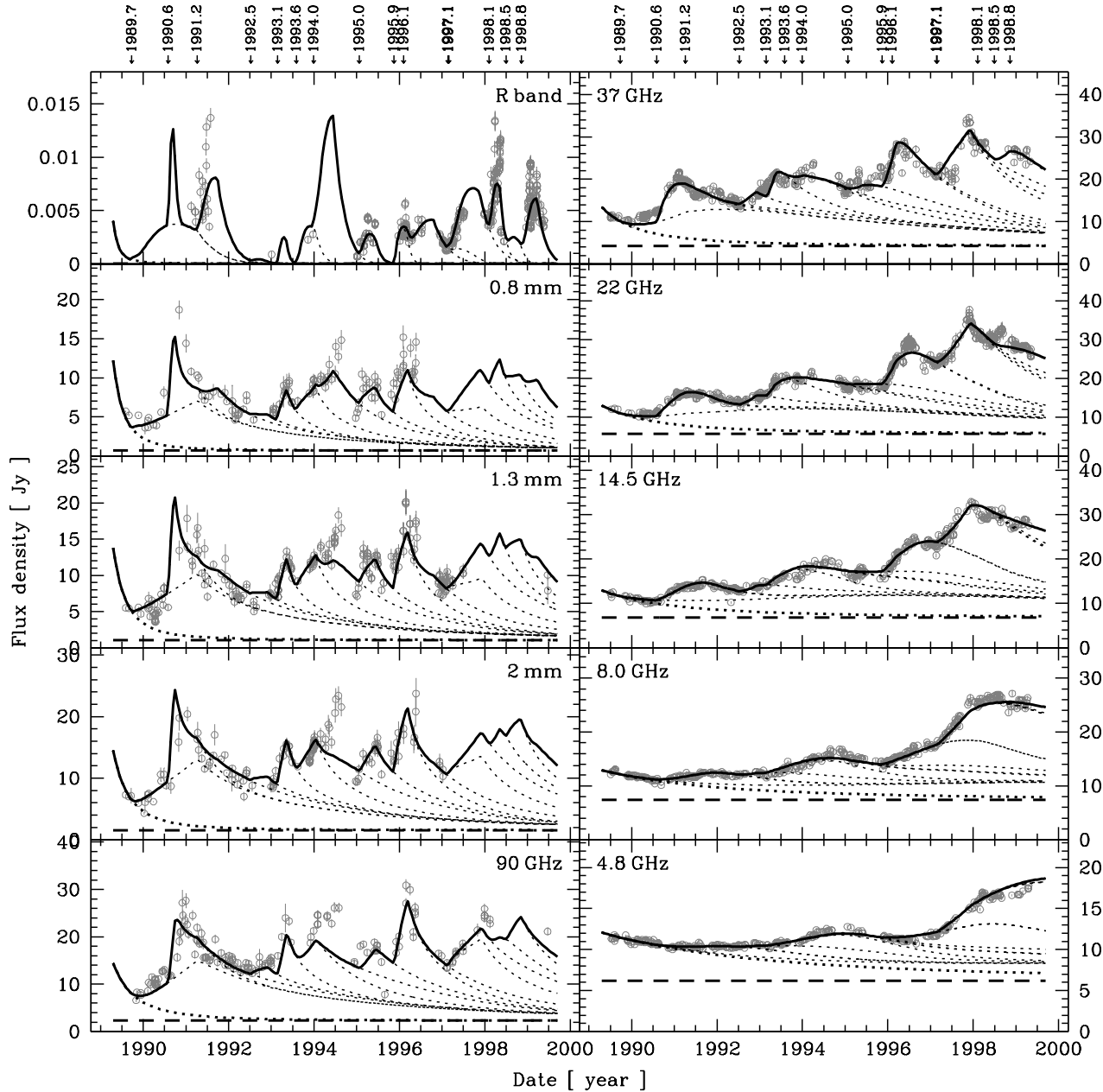}
\caption{Ten out of nineteen radio-to-optical lightcurves of
3C~279. The points indicate the observed flux density and the solid line our
best fit which is a sum of the underlying jet (long dashed line) and the
fifteen outbursts (dotted lines). The first dotted line represents
the global decay of all outbursts peaking before 1989.0}
\label{FigFinal}%
\end{figure*}

\section{Data}

To decompose the radio-to-optical lightcurves of 3C~279 we used data from five
radio and seven mm/submm wavelengths as well as four infrared and two optical
wavebands. The data sample extends from 1989.0 to 1999.5.

The 4.8, 8.0 and 14.5\,GHz data are from the University of Michigan Radio
Astronomy Observatory (UMRAO). Some of these data have not previously been
published.  The 22 and 37\,GHz data and some of the 90\,GHz data are from the
Mets\"ahovi Radio Observatory. These data were published by Ter\"asranta et
al. (1998, 2004).  We have also used 90 and 230\,GHz (1.3\,mm) data from the
Swedish-ESO Sub-millimeter Telescope (SEST). SEST data taken prior to 1994.5
were published by Tornikoski et al. (1996) while more recent data are
published here for the first time. From the Institut de Radio Astronomie
Millim\'etrique (IRAM) we have data at 90, 150 and 230 GHz (Steppe et al. 1993;
Reuter et al. 1997).  We also used data at 2.0, 1.3, 1.1, 0.85, 0.8, 0.45 and
0.35\,mm from the James Clerk Maxwell Telescope (JCMT). The 0.85\,mm dataset
was published by Robson et al. (2001) while data in the other wavebands are
either from Stevens et al. (1994) or are previously unpublished (post
1993). The dataset of Stevens et al. (1994) also includes the four infrared
wavebands used in this work.

The optical R- and V-band lightcurves are collected from the
literature, but also contain previously unpublished data. We have used
the data from Maraschi et al. (1994), Hartman et al. (1996), Villata
et al. (1997), Katajainen et al. (2000) and Hartman et al. (2001). The
previously unpublished data are from the Kungliga Vetenskapsakademien
(KVA) telescope and the Nordic Optical Telescope (NOT) as well as from
the observatories of Perugia and Torino.

\section{Model and Method}

In this paper we study the multifrequency lightcurves of 3C~279 in the
context of the shock model of Marscher \& Gear (1985). As the shock propagates
downstream in the relativistic jet it evolves through three stages. In
the initial (growth) stage, when inverse Compton losses predominate, the synchrotron
self-absorption turnover frequency decreases and the turnover flux density
increases. In the second (plateau) stage, synchrotron losses dominate and the
turnover frequency decreases while the turnover flux density remains roughly
constant. During the third (decay) stage when adiabatic losses
dominate, both turnover frequency and turnover flux density decrease.

Bjornsson \& Aslaksen (2000) criticize one of the assumptions of Marscher \&
Gear (1985) which concerns the rise of the peak flux density during the initial
Compton stage.  Their correction of the expression for the energy density of
synchrotron photons implies a much shallower rise of peak flux density with
peak frequency at the beginning of the outburst. On the other hand, multiple
Compton scattering, a process not considered by Marscher \& Gear (1985), would
actually steepen their derived relation.  There is therefore no compelling
theoretical reason to use a modified version of the original Marscher \& Gear
(1985) model.

In T\"urler et al. (2000), a generalisation of this model was used to decompose
the multifrequency lightcurves of 3C~273. We adopt a similar methodology in
this work. The shape of the emission spectrum behind the shock front is assumed
to be that of a simple synchrotron spectrum (with electron energy distribution
$N(E)\propto KE^{-s}$) with two spectral breaks, namely the low-frequency break,
$\nu_{\mathrm{h}}$, below which $\alpha_{\mathrm{thick}}=5/2$ and the high
frequency break, $\nu_{\mathrm{b}}$, which steepens $\alpha_{\mathrm{thin}}$ from
$(1-s)/2$ to $-s/2$. To extend the decomposition into the infrared and
optical we introduce an additional high frequency exponential cut-off to the
synchrotron spectrum corresponding to emission from the most energetic
electrons. The sharp low- and high-frequency spectral breaks of T\"urler et
al. (2000) were also replaced by more realistic smooth transitions. As in the
modelling of the micro-quasar GRS~1915+105 (T\"urler et al. 2004), we
introduced a new parameter allowing the ratio
$\nu_{\mathrm{h}}/\nu_{\mathrm{m}}$ of the low-frequency spectral break,
$\nu_{\mathrm{h}}$, and the synchrotron self-absorption turnover,
$\nu_{\mathrm{m}}$, to vary with time.

Another significant change is the addition of an underlying jet component with
constant flux density. This component is assumed to be an inhomogeneous
synchrotron source with a spectrum defined by a high-frequency break fixed at
375\,GHz -- as suggested by the quiescent level studies of 3C~279 (Litchfield
et al. 1995) -- and by two free parameters of the fit defining the synchrotron
self-absorption flux density and frequency. To be consistent with the lowest
infrared measurements an exponential cut-off was also added to the underlying
jet spectrum at a somewhat arbitrary frequency of 2$\cdot10^5$\,GHz. We note,
however, that our conclusions are insensitive to these numerical values.

The specificity of the individual outbursts is modelled by varying the scale of
their evolution in flux density, frequency and time as done in T\"urler et
al. (1999). The more physical approach adopted by T\"urler et al. (2000) of
varying the onset values of $K$ (electron energy density normalization), $B$
(magnetic field strength) and $D$ (Doppler factor) was also tried, but was
found to limit the differences between outbursts too much to get an equally
good fit.

The model presented here assumes a constant Doppler factor $D\propto R^{-d}$
and a magnetic field $B\propto R^{-b}$ perpendicular to the jet axis as
suggested by VLBI polarization measurements (Lister et al. 1998, Marscher et
al. 2002). The exponents, $b$ and $d$, of the evolution of these quantities with
radius, $R$, of the jet are therefore fixed to $d=0$ and $b=1$. We tried varying
these values as well, but this did not result in significantly better fits.

The resulting model has a total of 77 free parameters to fit the 15 outbursts
identified in the lightcurves of 3C~279 from 1989 to 2000. Twelve parameters
are used to describe the shape and evolution of the synchrotron spectrum for an
average outburst. Three further parameters describe the initial flux density
decay while two are needed to describe the underlying jet component.  The
remaining 60 parameters are used to describe the start time and the specific
characteristics of the 15 different outbursts. The total number of degrees of
freedom is 3281.

\begin{figure*}
%\sidecaption \centering 
\includegraphics[width=12cm]{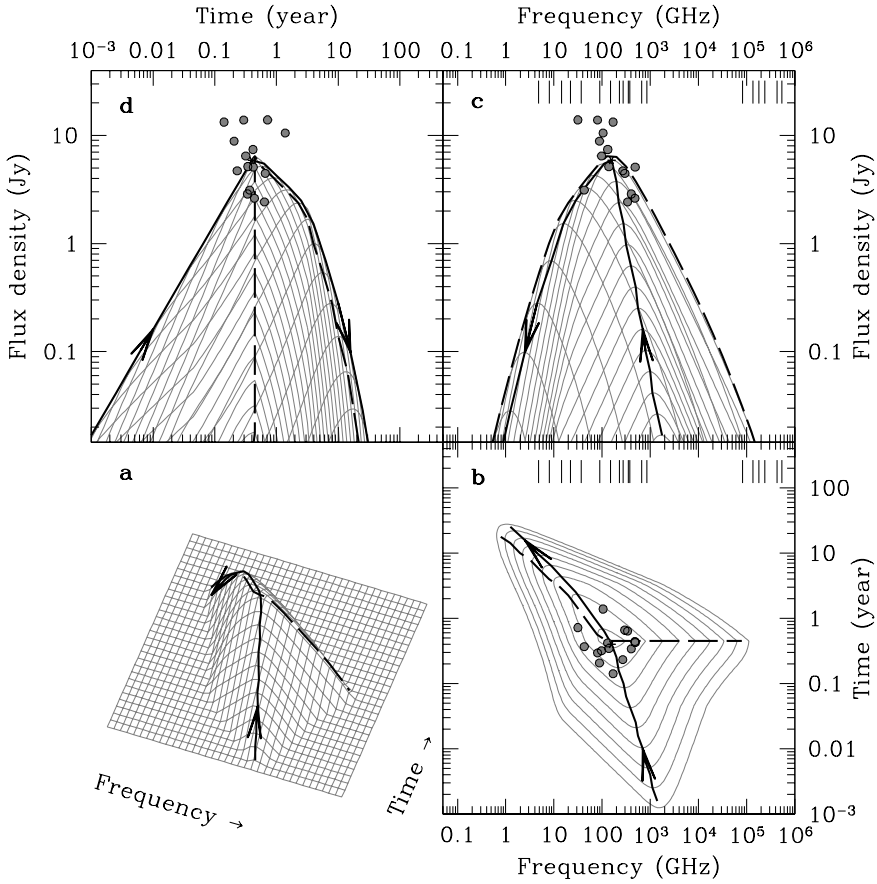}
\hfill
\parbox[b]{55mm}{
\caption{Logarithmic evolution of the synchrotron emission of an average
  outburst in 3C~279. The three dimensional (log $\nu$, log $S$, log $t$)
  representation is shown in (\textbf{a}), while the Cartesian projections of
  this surface are shown in (\textbf{b}), (\textbf{c}) and (\textbf{d}).
  \textbf{b}) Frequency versus time representation with contours starting at a
  flux density of 0.02 Jy and spaced by 0.3 dex. \textbf{c}) The synchrotron
  spectra at different times spaced by 0.2 dex. \textbf{d}) The lightcurves at
  different frequencies spaced by 0.2 dex. The thick solid line with arrows
  traces the time evolution of the spectral turnover, whereas the dashed line
  connects the peak of the lightcurves at different frequencies. The filled
  circles refer to the specific characteristics of the outbursts (see Section
  4.2) and the vertical bars in (\textbf{b}) and (\textbf{c}) show the
  frequency coverage of the 19 lightcurves.}
}
\label{FigEvolution}%
\end{figure*}

\section{Results}
The best-fit decomposition of the radio-to-optical lightcurves into a series of
15 self-similar outbursts is shown in Fig.~1. It has a reduced $\chi^2$ value
of 8.72. This relatively high value can be considered as acceptable here
keeping in mind that the aim of the modelling is to derive the average
properties of a typical outburst and that the specificity of each outburst is
only modelled very crudely in order to minimise the number of free
parameters. Furthermore, the analytical shock-in-jet model itself is of
necessity a simplistic idealization of the complex emission of actual jets
(G\'omez 2005) and thus a perfect match of the model to the data cannot be
expected. At the lowest frequencies, the fit describes the smooth shape of the
lightcurves very well. At 90\,GHz, the general flux density level of the fit is
good, but during the biggest outbursts (in 1991, 1993.5, 1994 and 1996.5) the
model flux density remains below the observations. The same tendency also
applies to all mm-band lightcurves and to the optical R- and V-band
lightcurves. Part of this fit (until 1994.0) has been published in Lindfors et
al. (2005).

\subsection{Average evolution of the synchrotron spectrum}

The average evolution of an outburst in 3C~279, as defined by 12 fit
parameters, is shown in Fig.~2. In Fig.~2c we see the average outburst
spectra. It peaks at $\sim$150\,GHz, which is the frequency at which the
transition from the Compton stage to the synchrotron stage takes place. The peak
flux density is 6.6 Jy. The second stage transition in the average spectra is
at $\sim 20$\,GHz and its flux density is 2.2 Jy.

The overall shape of the evolution is quite different from the one derived for
3C~273 (T\"urler et al. 2000). The most obvious difference is the decreasing
flux density during the second stage of the evolution which is quite different
from the typical plateau seen in 3C~273. In the model, this difference can be
ascribed to the high value of the parameter $k$ defining the decrease of the
normalization $K\propto R^{-k}$ of the electron energy distribution
$N(E)\propto K\,E^{-s}$ with jet opening radius, $R$. The derived value of $k$
is $4.0$ which is far above the value of $k_{\mathrm{ad}}=2(s+2)/3=2.81$ (with
$s=2.25$) predicted for an adiabatic jet flow. The original Marscher \& Gear
model adopted only adiabatic compression to heat the electrons in the
shock. While this seems to be a good assumption for 3C~273 (k=3.03), our high
$k$-value found for 3C~279 suggests that the emitting electrons are subject to
important non-adiabatic cooling processes within the jet. This might be due, at
least in part, to synchrotron and inverse Compton radiative losses in the
undisturbed underlying jet.

As mentioned in T\"urler et al. (2000), the three parameters $r$, $k$ and $d$
are difficult to constrain uniquely by the fit. In particular, the effect of
$k>k_{ad}$ and of $d>0$ are similar, both producing a decreasing flux density
during the synchrotron stage of the evolution. The high value of $k$ might
therefore also indicate that our assumption of $d=0$ is not valid and thus
suggest that the Doppler factor tends to decrease during the evolution of the
shock. This could be due to a decelerating jet flow or to a geometry in which
the jet bends away from the line of sight (see Jorstad et al. 2004).

Another difference is that the flattening of the spectral index by
$\Delta\alpha_{thin}=+0.5$ (predicted to occur at the synchrotron-to-adiabatic
stage transition) is found to be much less abrupt for 3C~279 than for 3C~273,
and for 3C~279 it starts much earlier during the initial Compton stage. This
can be seen by the position of the start of the flattening ($t_\mathrm{f}$) in
Fig.~3, which is much earlier than found for 3C~273 (see Fig.~4. of T\"urler et
al. (2000)). A similar behaviour was also found for GRS~1915+105 (T\"urler et
al. 2004).

Apart from $k$, the values of two other physical parameters of the jet are
found to be similar to those derived for 3C~273 (T\"urler et al. 2000). The
index, $s$, of the electron energy distribution is found here to be $s=2.25$
($s=2.05$ for 3C~273) and the opening radius $R\propto L^r$ of the jet with
distance, $L$, along the flow is found to be slightly non-linear ($r=0.78$)
indicating that, as for 3C~273 ($r=0.8$), the jet opening angle is slowly
decreasing with distance suggesting that some jet collimation process is at
work.

We also note, as was found for GRS~1915+105 (T\"urler et al. 2004), that there is a
tendency for the synchrotron emission of the outbursts to start out inhomogeneous
but to become homogeneous as the shock evolves with time. This is manifested as
a decreasing $\nu_{\mathrm{h}}/\nu_{\mathrm{m}}$ ratio
%of the low-frequency
%spectral break, $\nu_{\mathrm{h}}$, and the spectral turnover,
%$\nu_{\mathrm{m}}$. 
which affects the shape of the spectrum at frequencies below
the turnover as can be seen in Fig.~2c.

Another difference is that in 3C~273 the outbursts are very short-lived at
higher frequencies, while in 3C~279 at all frequencies above $\sim$150\,GHz it
takes $\sim$ 0.45 years to reach the peak flux density. Since the rise time at all
such high frequencies is the same these outbursts peak
simultaneously. This is shown in Fig.~3. This result, together with the finding that
the plateau stage is a bit shorter than in 3C~273, might suggest that in 3C~279
the magnetic field energy density is relatively lower than the electron energy
density.

\begin{figure}
\centering \includegraphics[angle=0,width=0.48\textwidth]{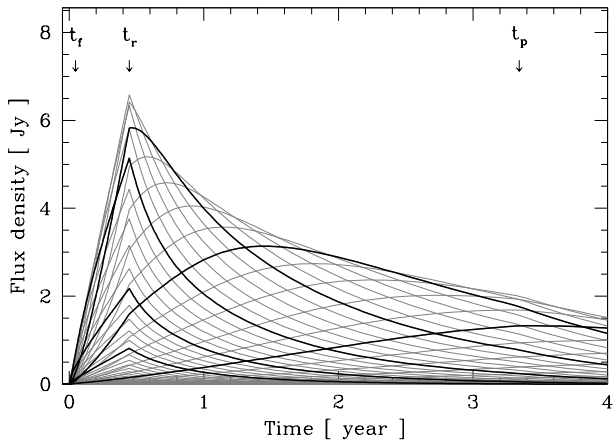}
\caption{Model lightcurves of the average outburst in 3C~279 at different
  frequencies spaced by 0.1 dex (grey lines). The six highlighted lightcurves
  are at frequencies, $\nu$, defined by log($\nu$/GHz)= 3.5, 2.5,...,1.0, in
  order of increasing timescales. t$_\mathrm{r}$ and t$_\mathrm{p}$ refer to
  transitions from Compton to synchrotron stage (t$_\mathrm{r}=$0.45 years) and
  from synchrotron to adiabatic stage (t$_\mathrm{p}=$3.34 years)
  respectively. t$_\mathrm{f}$ refers to the time when the flattening of the
  spectral index starts (t$_\mathrm{f}=$0.046 years).}
\label{FigEvoLC}%
\end{figure} 

3C~279 is an archetypical blazar while 3C~273 is not a true member of the
blazar class. Most conspicuously, it is not rapidly variable in the optical
band and has low optical polarization. However, it exhibits blazar-like optical
behaviour at low levels, diluted by thermal radiation (Impey et al. 1989,
Valtaoja et al. 1990, Valtaoja et al. 1991) and has therefore been dubbed a
mini-blazar by Impey et al. (1989). As expected within the context of the
unification scheme, 3C~273 has a rather large viewing angle, estimated to be
around 10 degrees (L\"ahteenm\"aki \& Valtaoja 1999, Savolainen et
al. 2006). This is smaller than for ordinary quasars, but larger than for
blazars; for 3C~279 the estimated viewing angle is around 2 degrees (Lindfors
et al. 2005 and references therein).

In the simplest unification models, orientation is the only parameter. One
might therefore expect that the borderline blazar 3C~273 would appear similar
to 3C~279 if the viewing angle were decreased. However, some of the differences
we find between 3C~273 and 3C~279 (non-adiabatic jet flow in 3C~279, adiabatic
in 3C~273; magnetic energy density lower than electron energy density in 3C~279
whereas in 3C~273 there is little difference) seem intrinsic to the sources and
cannot be explained by differences in viewing angles alone. However, we note
that variations in the viewing angle as a function of time/radial distance
could play a role in the observed differences between these two sources.
  
\subsection{Characteristics of Different Outbursts}
The evolution of each of the 15 individual outbursts is allowed to deviate in
scale from the average outburst. For this we need three parameters
corresponding to the logarithmic shifts along the three axes of Fig.~2, namely
time, $\Delta\log t$, frequency, $\Delta\log \nu$ and flux density, $\Delta\log
S$. The values obtained for these shifts are given in Table 1 and their
distribution is shown graphically with the filled circles in Fig.~2.

\begin{table}
\begin{center}
\begin{tabular}{l l r r r}\hline \hline
Shock nr.&$T_0$&$\Delta$ log$S$&$\Delta$ log$\nu$&$\Delta$ log$t$\\ \hline
1&1989.72&0.25&-0.18&0.54\\ 
2&1990.57&0.35&0.03&-0.45\\ 
3&1991.25&-0.35&0.48&0.04\\ 
4&1992.51&-0.28&-0.57&-0.03\\ 
5&1993.14&0.18&-0.25&-0.29\\ 
6&1993.57&0.10&-0.08& 0.02\\ 
7&1993.98&-0.06&0.49&0.03 \\ 
8&1995.05&-0.06&-0.06&-0.06\\ 
9&1995.86&0.37&-0.29&-0.13\\ 
10&1996.09&-0.38&0.33&0.20\\ 
11&1997.12&0.37&-0.70&0.26\\ 
12&1997.14&-0.12&0.27& 0.22\\ 
13&1998.09&-0.10&0.23&-0.24\\ 
14&1998.48&0.04&-0.21&-0.10\\ 
15&1998.85&-0.31&0.41&-0.07\\\hline 
\end{tabular}
\caption{The start times of the outbursts ($T_0$) and their logarithmic
  shifts from the average shock evolution in flux density ($\Delta$ log $S$),
  frequency ($\Delta$ log $\nu$, negative if the outburst peaks at lower
  frequencies than the average outburst) and time scale ($\Delta$ log $t$,
  negative if the outburst evolution is faster than the evolution of the
  average outburst).}
\end{center}
\end{table}

We attempted to understand the physical origin of the differences seen between
the outbursts by looking for correlations between $\Delta \log S$, $\Delta \log
\nu$ and $\Delta \log t$. Contrary to results for 3C~273 (T\"urler et
al. 1999) we find no correlation between any of the shifts, and Fig.~2 shows
clearly that the shifts do not align with any axis. It is also clear that the
outbursts do not differ mainly in amplitude as found for GRS~1915+105 (T\"urler
et al. 2004) but also in peaking frequency and duration.

To establish that our decomposition is not purely mathematical but corresponds
to some physical reality, we compare the outbursts suggested by our fit to
observed VLBI components. 3C~279 has been monitored with VLBI since the 1980s,
and for the period considered in this paper the VLBI data coverage is good. In
Table~2, the start times ($T_0$) of the outbursts, as derived from our
modelling, are compared with the extrapolated zero-epochs of the VLBI
components (Wehrle et al. 2001, hereafter W01; Jorstad et al. 2004, hereafter
J04). We also compare the observed flux densities of these knots (from W01 and
Jorstad et al. 2005, hereafter J05) to the model flux densities of the
outbursts (at the same date and frequency as the VLBI observations). We adopt
the naming scheme of the VLBI knots from W01 (components C4 to C7a) and J04
(components C8 to C13). We note that there is a difference in identification of
component C9 between these two papers and we choose to adopt the identification
from the latter.

\begin{table*}
%\begin{tiny}
\begin{center}
\begin{tabular}{l l l l l l l l l}\hline \hline
nr.&$T_0$& Knot &$T_0$(knot)& F$_{\mathrm{exp}}$& F$_{\mathrm{obs}}$&time&freq.&Rise Time\\ 
&&&&[Jy]&[Jy]&of obs.& [GHz]&at 43GHz\\ \hline
&&C4&1984.68$\pm^{0.27}_{0.29}$&&1.42& 1991.47&22& \\ 
1&1989.72&C5&?&3.4&2.18&1991.47&22&2.38 \\ 
2&1990.57&C5a&1990.88$\pm^{0.28}_{0.39}$&5.3 &1.82& 1991.47& 22&0.39 \\ 
3&1991.25&&&0.01 &&1991.47&22&3.19\\ 
4&1992.51&C6&1992.09$\pm^{0.14}_{0.17}$&2.2 &2.52&1992.86&22&0.41\\ 
5&1993.14&C7&1993.26$\pm^{0.13}_{0.14}$&4.3 &3.80&1994.17&22&0.30\\ 
6&1993.57&&&1.85&&1994.17&22&0.90\\ 
7&1993.98&&&0.35&&1996.02&22&3.11\\ 
8&1995.05&C7a&1994.67$\pm^{0.04}_{0.05}$&2.0 &2.86$^{*}$&1996.02&22&0.78\\ 
9&1995.86&C8&1995.70$\pm0.1$&2.3&&1996.02&22&0.40\\ 
10&1996.09&&&0.5&&1997.47&43&3.57\\ 
11&1997.12& C9&1996.89$\pm0.1$&10.7&2.959&1998.23&43&0.81\\ 
12&1997.14&C10&1997.42$\pm0.1$&1.0&3.295&1998.23&43&3.21\\ 
13&1998.09&C11&1997.59$\pm0.1$&0.4&4.305&1998.23&43&1.02\\ 
14&1998.48&C12&1998.50$\pm0.1$&5.2&1.312&1998.94&43&0.51\\ 
15&1998.85&C13&1998.98$\pm0.1$&0.05&2.295&1998.94&43&2.16\\ \hline 
\end{tabular}
\end{center} \begin{tabular}{l}
$^*$sum of C7a and C8
\end{tabular}
%\end{tiny}
\caption{Comparison of the outbursts found here with observed VLBI
  components. The start times, $T_0$, of the outbursts are compared to the
  zero-separation epochs, $T_0$(knot), of VLBI components derived by Wehrle et
  al.  (2001) and Jorstad et al. (2004). We also compare the extrapolated flux
  density, F$_{\mathrm{exp}}$, of the outbursts at the time of the VLBI maps to
  the observed VLBI flux density, F$_{\mathrm{obs}}$ (Wehrle et al. 2001;
  Jorstad et al. 2005). The ninth column gives the rise times of the outbursts
  at 43\,GHz.}
\end{table*}

Throughout the 1990s the VLBI maps showed two permanent features: C4 and C5. C4
is a very long-lived feature in 3C~279. The flux density of this component has
not decreased steadily with time but has dimmed or brightened on several
occasions. This component contributes significantly to the initial flux density
decay (which is a sum of all outbursts peaking before 1989.0) but our model
cannot account for its brightenings. For C5 there is no zero-separation epoch
available because its fitted motion is consistent with zero speed. The flux
density of this component was falling throughout the 1990s and it disappeared
in 1998. We suggest that the first, very long-lived outburst in our fit is
related to this stationary component and is not caused by a moving knot.

We identify the other components by looking at the closest match between the
zero-separation epoch of the VLBI component and the start time of the
outburst. We find that for most of the components the match is rather good,
although in most cases $T_0$ is outside the error-bars of the extrapolated VLBI
zero-separation time. There are large mismatches between the observed VLBI
component flux densities (from W01 and J05) and the flux densities derived for
our model components. The model outbursts have systematically higher flux
densities than the real VLBI components.  On the other hand, the flux density
of the underlying jet component in our model is lower than the flux density of
the VLBI core at any given epoch. This suggests that the underlying jet
component is underestimated in our decomposition, while the flux density of the
outbursts is overestimated during their decay stage.

We can see from Table~2 that there are four outbursts without a corresponding
VLBI component. The third and the seventh outbursts are high-frequency peaking
outbursts (see Table~1) which are probably too weak to be seen in the maps by
the time the decaying shocks are emitting at VLBI frequencies. On the other
hand, the sixth outburst should be seen in the VLBI map but when we look at
the 22 and 37\,GHz lightcurves (those closest to the VLBI frequencies) we can
identify the fifth and the sixth outburst as a single event. The millimeter
lightcurves cannot be fitted with one outburst and we therefore suggest that in
higher frequency VLBI we might see two components. Another possible explanation
is that the fifth outburst is just a core flare (it has rather short duration,
see Table~1) and that the VLBI component should be recognized with the sixth
outburst. The VLBI observations of W01 show that the core flux density
increased during 1993, thus supporting the latter scenario. The long-lived
tenth outburst, which is rather faint at low frequencies, is likely to be
related to the ninth outburst and to the exceptionally bright VLBI component
C8. In fact, it might be the case that we see the tenth outburst of the fit
because of the brightening of the component C8 seen in VLBI maps by W01 and
J04. The mismatch between the flux densities of the outbursts and VLBI
components is very severe during 1997 (eleventh, twelfth and thirteenth
outbursts). VLBI observations from this period suggest that the direction of the
jet ``nozzle'' changed (J04). Our simple model cannot reproduce this behaviour
and therefore the discrepancy in component flux densities is bigger than for
other outbursts.

The fifteenth outburst peaks at a high frequency, similar to the third and the
seventh outburst, therefore it has a very low flux density at VLBI frequencies
and should not be visible in the maps. It is therefore likely to be unrelated
to the VLBI component C13, which then has no outburst counterpart in our
fit. This is probably because the lightcurves we use in our fit end soon after
the zero epoch of C13. As the last flux density points at 22 and 37\,GHz are
decreasing we cannot fit an additional outburst there even if one was
beginning at that time.

To make a further comparison of our model fit to VLBI observations we produced
artificial time-series of VLBI maps of 3C~279 at 22\,GHz. These five maps
describing the evolution of our model jet are shown in Fig.~4. The flux density
of the underlying jet component of our fit is fixed to the core and remains
constant at a level of 5.8~Jy. The components of the map correspond to model
outbursts of our fit and we assume them to move with a constant apparent
velocity of 5$c$ (W01 found that components have apparent velocities of 4.8$c$
to 7.5$c$). The maps are scaled to the distance of 3C~279 assuming $H_0=70$ km
s$^{-1}$ Mpc$^{-1}$, and their angular resolution is set to 0.2\,mas as in
Fig.~3 of W01 to ease the comparison.

At first sight, the simulated maps look very different to the observed
ones. This is, however, mainly due to the fact that the components starting
before 1989 could not be included in the simulated maps and explains the growth
of the simulated jet as well as the absence of component C4. Another clear
difference is the bright outer blob in the simulated maps, which corresponds to
the first, very long-lived outburst. As stated above, this is likely not a
real propagating shock wave, but accounts for the flux density of the
stationary component C5 and, maybe also to some extent, to the strong
long-lived and re-brightening component C4. Apart from this, we note a
predominance of the core emission in the observed VLBI maps compared to our
simulated maps. This illustrates the fact that the flux density of the core in
the VLBI maps of W01 is higher at all times than the flux density of our
underlying jet component, and it varies from 8.8 to 21.1 Jy.  W01 found that
flux density variations of the core are largely responsible for the total flux
density variability observed by single-dish antennae. The core brightenings are
often followed by the ejection of superluminal components. The model maps
presented here suggest that during the initial stage of the model outbursts it
is not possible to resolve them from the VLBI core; they are first seen as a
brightening of the core. In the later stages, the longer lasting outbursts are
seen as VLBI components while short-lived components might only be seen as
core-flares due to the limited resolution of the instrumentation.

Using the shifts given in Table~1, we can calculate the rise time of each
outburst at 43\,GHz, defined as the duration from the beginning of the flare to
the time when the maximum flux density at 43\,GHz is reached. From the values
given in Table~2, we note that most of the outbursts peak within 1 year and
they have a maximum flux density amplitude at rather low frequencies. Using the
same assumptions as above, the proper motion is 0.16 mas/yr, so most of the
flares peak while they are still within the core (even if we would assume a
resolution of $>0.1$ mas like in J05).  When the component is resolved in VLBI
maps, the flux density is already decaying but the time elapsed since the
shock's maximum stage is short and the component is still well detectable with
VLBI. On the other hand, the outbursts with a very long rise time at 43\,GHz
are associated with shocks peaking at high frequencies which are heavily
self-absorbed at VLBI frequencies.  After several years, when the
self-absorption turnover reaches 43\,GHz, the component flux density has
already decreased appreciably and often remains below the VLBI detection
limits.

\begin{figure}
\centering \includegraphics[angle=0,width=0.48\textwidth]{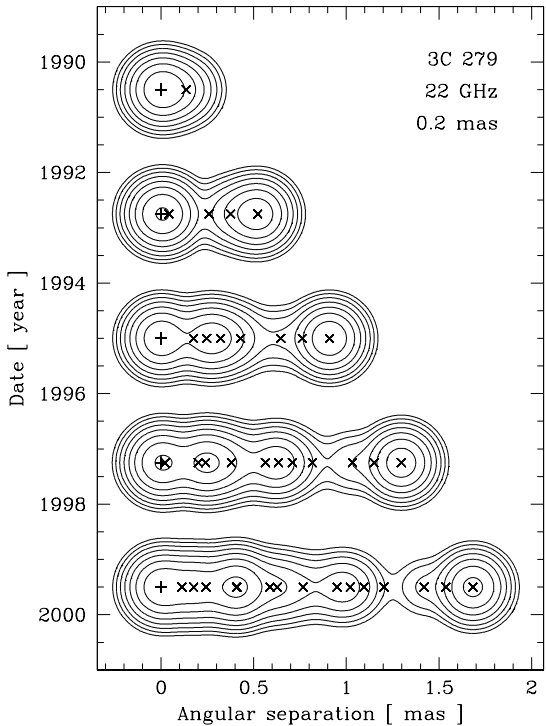}
\caption{A time sequence of five model VLBI maps of 3C~279 at 22\,GHz between
1990 and 2000. As in Fig.~3 of W01, the map resolution is 0.2 mas, with a
lowest contour at 50 mJy beam$^{-1}$ and subsequent contours increasing by
factors of 2. Each component (marked with $\times$) is assumed to move with a
constant apparent velocity of 5$c$ with respect to the core of the jet (marked
with $+$). An animation of this figure is shown at:
http://isdc.unige.ch/$\sim$turler/jets/.}
\label{FigVLBI}%
\end{figure}

We conclude that the outbursts of our fit correspond satisfactorily to real
observed components in the jet. This is in accordance with previous studies:
T\"urler et al. (1999) found that their model outbursts correspond to VLBI
components in 3C~273, and Savolainen et al. (2002) studied a sample of 27 AGN
and found a clear connection between the observed VLBI components and the total
flux density outbursts in 22 and 37\,GHz radio lightcurves.

\subsection{Comparison with EGRET gamma-ray states}
As 3C~279 is the most frequently observed EGRET blazar it is very interesting
to take a closer look at the components present at different EGRET epochs. It
is reasonable to assume that most of the gamma-ray emission will originate in
the newest shock component since, during the early stages of shock evolution,
the dominant cooling mechanism is inverse Compton losses. This shock is usually
also the dominant component in the submm to optical range of the spectrum (see
below). Therefore we calculated the distances of these components along the jet
with:

\begin{equation}
L\approx D^2\cdot c(1+z)^{-1}\Delta t_{\mathrm{obs}}
\end{equation}
where z=0.538 for 3C~279 and for simplicity we assume $D=10$ for all shock
components, in accordance with Lindfors et al. (2005). $\Delta
t_{\mathrm{obs}}$ is the difference between the onset time of the outburst
($T_0$) and the time of the EGRET observation. The results are shown in
Table~3.

\begin{table}
\begin{center}
\begin{tabular}{l l l l l}\hline \hline
Epoch& Shock nr.&$\Delta t_{obs}$& gamma-ray state& $L$\,[pc]\\ \hline
P5b (1996.092)&10&0.006&very large flare&0.12\\
P8 (1999.070)&15&0.195&high&3.88\\
P5a (1996.063)&9&0.206&high&4.10\\
P1 (1991.47)&3&0.225&high&4.48\\ 
P3a (1993.858)&6&0.288&moderate&5.74\\ 
P6b (1997.470)&12&0.328&moderate&6.53\\ 
P3b (1993.979)&6&0.409&moderate&8.15\\
P2 (1993.004)&4&0.49 &low&9.76\\ 
P6a (1997.010)&10&0.924&low&18.41\\
P4 (1994.970)&7&0.99&low&19.72\\ \hline  
\end{tabular}
\caption{The EGRET epochs ordered by increasing $\Delta t_{\mathrm{obs}}$
  [years], the time interval since the start of the last outburst as derived
  here. This order is compared to the gamma-ray state adopted from Hartman et
  al. (2001). Also an estimate of the distance, $L$, of these shock components
  from the apex of the jet is given (see text).}
\end{center}
\end{table}

We note that all but one (outburst 10 at epoch 5b) of the shock components are
far beyond the broad-line region (BLR) boundary of 0.4 pc (Kaspi et al.2000) at
the time of the EGRET observation. Therefore, if the gamma-ray emission is
indeed related to shocks travelling in the jet then mechanisms based on 
external photon fields reprocessed in the BLR -- actually most of the External
Compton mechanisms suggested up to date (e.g. Sikora et al. 1994; Hartman et
al. 2001, and references therein) -- are ruled out for all other epochs but 5b,
when a very large flare in gamma-rays was observed.

There is a clear pattern in Table~3 when we look at the $\Delta
t_{\mathrm{obs}}$ and compare them to the gamma-ray state. The high states (and
the very large flare in epoch 5b) correspond to the shortest $\Delta
t_{\mathrm{obs}}$, while the low states correspond to the longest $\Delta
t_{\mathrm{obs}}$ and the moderate states fall neatly between these two
cases. This pattern clearly suggests that the gamma-ray emission is indeed
connected to the shocks travelling in the jet and that the highest gamma-ray
fluxes are observed when there is an early stage shock present in the jet.

To gain confidence in the reality of this pattern, we also looked for a
correlation with the next outburst beginning after, rather than before, a
gamma-ray epoch. But in this case, there is apparently no pattern between the
gamma-ray state and the time delay between the gamma-ray epoch and the
beginning of the outburst.

It is instructive to compute the probability that the 10 $\Delta
t_{\mathrm{obs}}$ are grouped as they appear in Table 3 by chance, assuming no
real correlation exists between $\Delta t_{\mathrm{obs}}$ and gamma-ray
state. We simulate 100,000 sequences of the 10 $\Delta t_{\mathrm{obs}}$ and
repeated the simulation 1000 times. We found that in average the $\Delta
t_{\mathrm{obs}}$ are ordered as in Table~3 24.45 times per simulation, which
corresponds to a probability of 0.024\% or $3.7\sigma$ away from the excepted
value (assuming a normal distribution).

In the simulations we assume that $\Delta t_{\mathrm{obs}}$ and the gamma-ray
state are well defined quantities with no uncertainties. For low-peaking
outbursts, the error in the determination of $T_0$ (and therefore $\Delta
t_{\mathrm{obs}}$) is rather small because the radio lightcurves are very well
sampled. For high-peaking outbursts the uncertainty is larger so, for example,
we cannot exclude the possibility that outburst 10 would start 0.01 years later
and the dominating outburst at epoch 5b would be outburst 9 with $\Delta
t_{\mathrm{obs}}=0.235$. This would still not destroy the pattern seen in
Table~3, but demonstrates that although the observed correlation has a high
statistical significance, it should still be considered suggestive rather than
definite.

It has already been established statistically for a sample of sources that
gamma-ray flares are connected with growing shock components propagating in the
jet (Jorstad et al. 2001, L\"ahteenm\"aki \& Valtaoja 2003). Our finding
supports this scenario and suggests additionally that there is a dependence
between the gamma-ray flux and the distance of the shock component from the
apex of the jet. We emphasize the need of further studies on this issue when
future experiments provide gamma-ray lightcurves with better time
coverage.

\begin{figure*}
%\sidecaption
%\centering 
\includegraphics[width=12cm]{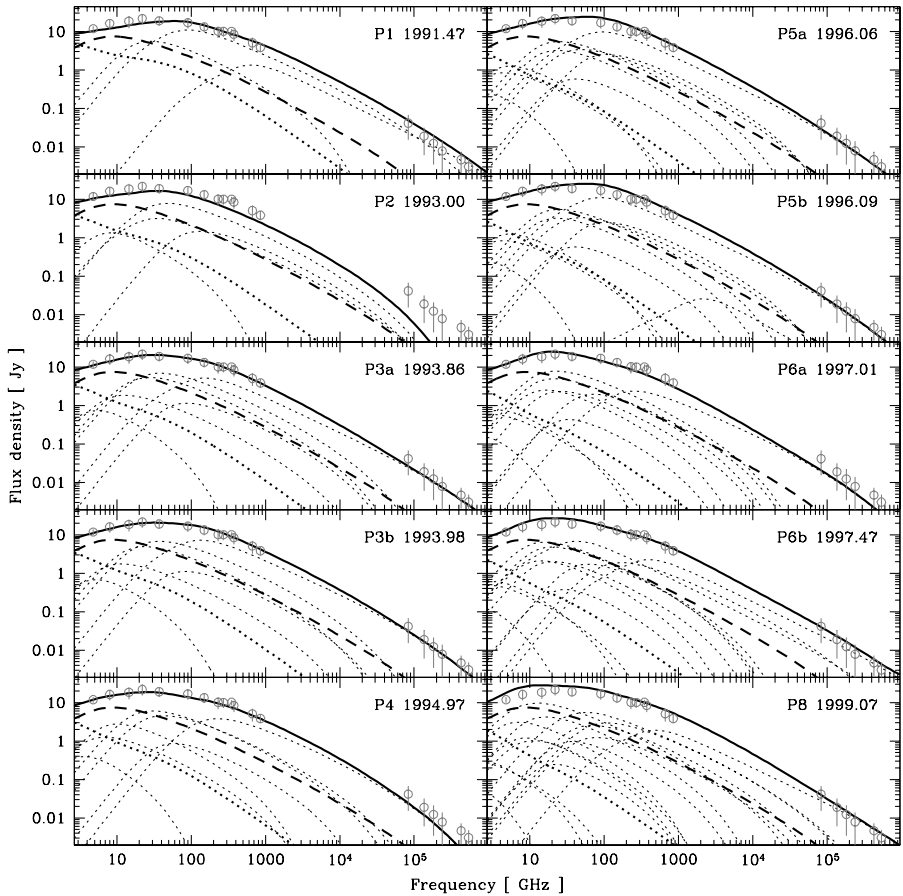}
\hfill
\parbox[b]{55mm}{
\caption{The synchrotron spectra of 3C~279 at 10 EGRET epochs. The points
indicate the observed mean flux density levels. The solid line is the total
model spectrum at each epoch which is the sum of the spectra of the underlying
jet component (dashed line), the overall spectrum of the outbursts peaking
before 1989.0 (thick dotted line) and the individual outbursts (dotted lines).}
}
\label{FigFinal}%
\end{figure*}

To discuss further the link between gamma-ray emission and synchrotron
outbursts we show in Fig.~5 the low-frequency spectral decomposition at each
EGRET epoch as derived from our lightcurve fitting. For each epoch, the total
model spectrum and its decomposition into individual outbursts is compared to
the average spectrum of 3C~279 (points). We see that at two of the high
gamma-ray state epochs (P1 and P8) the model spectrum in the optical range is
clearly above the average observed flux density levels, but at two other epochs
(5a and 5b) the model spectrum fits very well the average optical flux
density. At these epochs it is instead the mm/submm flux density which is above
the average, which is not seen in epochs P1 and P8. This is easy to understand
by looking at the dominant shock components at these epochs: at P1 and P8 the
last outburst is high-frequency peaking, while the ninth outburst at epochs 5a
and 5b peaks in the millimeter range.

There are four epochs when the fitted optical spectrum is below the observed
mean flux density level. Three of these epochs correspond to a low gamma-ray
state and the fourth to a moderate gamma-ray flux (epoch P6b). There are also
two epochs (P2 and P6a) corresponding to low gamma-ray states and for which the
millimeter to submillimeter spectrum is below the observed mean spectral energy
distribution. Finally, we note that the centimeter spectrum is dominated by the
underlying jet component at all epochs, and that the fitted spectrum is below
the observed one at the first two epochs but above it at the last three epochs.

In summary, the pattern we see in Fig.~5 is in accordance with and strengthens
the hypothesis that gamma-ray emission originates in young, growing shocks
which are dominant in the mm-to-optical regime at the time of high gamma-ray
states.

\section{Conclusions}
We have studied synchrotron flaring in the jet of 3C~279 by decomposing the
multifrequency lightcurves into a series of outbursts. The analysis is
performed within the generally accepted shock-in-jet scenario, with the
evolution of the model outbursts following the Marscher \& Gear (1985)
prescription. Similar methods have been used previously to study the quasar
3C~273 (T\"urler et al. 1999, 2000) and the micro-quasar GRS~1915+105. For
3C~279 we find:

1. The lightcurve decomposition supports the Marscher \& Gear model. The smooth
  low-frequency lightcurves are especially well reproduced by the model. In the
  high-frequency (millimeter to optical) domain, the sharp peaks of the
  observed lightcurves are less well reproduced by the model lightcurves during
  the strongest outbursts. Although the model used here makes a lot of
  assumptions and simplifications which might not hold in detail, it is still
  clear that there is some mechanism at work which is not well described by the
  Marscher \& Gear (1985) model, and which produces the sharp peaks of
  outbursts at high frequencies. This is in accordance with previous findings
  (e.g. Valtaoja et al. 1999): both the theoretical and numerical jet models
  always produce smooth peaks in model lightcurves while the observed outbursts
  have exponential rises, sharp peaks and exponential decays.

2. The outburst evolution in 3C~279 follows the characteristic three stage
  pattern but instead of the synchrotron plateau stage seen in 3C~273, the
  flux density is already decreasing during the second stage. This might be
  explained by a highly non-adiabatic jet flow, where radiative losses play a
  significant role. We also find that the lightcurve of an average outburst is
  different for 3C~273 and 3C~279. The differences found between these two
  sources appear to be mainly intrinsic and not due to different
  orientations. The borderline blazar 3C~273 is thus not simply 3C~279 seen
  from a larger viewing angle.

3. We find a rather good agreement between the zero-separation epoch of VLBI
  components and the start times of model outbursts. However, some
  high-frequency peaking outbursts do not have a corresponding VLBI component
  which can be explained by the long delay before their spectral turnover
  reaches VLBI frequencies.

4. When comparing the gamma-ray state at different epochs from Hartman et
   al. (2001) to the time elapsed since the start of the last synchrotron
   outburst we find that the shortest time intervals (corresponding to the
   shortest distances from the apex of the jet to the shock) are seen during
   the highest gamma-ray states, and the longest ones during the lowest gamma-ray
   states. This supports previous findings (e.g. Valtaoja \& Ter\"asranta 1996;
   Jorstad et al. 2001; L\"ahteenm\"aki \& Valtaoja 2003) that the gamma-ray
   flares are connected to shocks propagating in the jet. Although during the
   highest states the shock components are rather new, they are still well
   outside the outer limits of the broad-line region. This suggests that
   external Compton models based on the soft-photon field originating from the
   BLR are unlikely to be able to explain the gamma-ray flares in this source.
  
\begin{acknowledgements}
This research has been supported by the Academy of Finland grants 74886 and
80450 and Jenny and Antti Wihuri Foundation. This work was partly supported by
the European Community's Human Potential Programme under contract
HPRN-CT-2002-00321.  We also thank the referee for constructive criticism of
the earlier version of this paper.
\end{acknowledgements}

\end{document}